\documentclass[5p]{elsarticle}
\usepackage[utf8]{inputenc}
\usepackage{amssymb}
\usepackage{amsmath}
\usepackage{hyperref}
\begin{document}
\journal{Journal of Subatomic Particles and Cosmology}
\begin{frontmatter}
\title{The QCD phase diagram, universal scaling, and Lee-Yang zeros}
\author[1]{Christian Schmidt}
\ead{schmidt@physik.uni-bielefeld.de}
\affiliation[1]{organization={Universit\unexpanded{\"a}t Bielefeld},%
addressline={Fakult\unexpanded{ä}t f\unexpanded{ü}r Physik}, 
postcode={D-33615},
city={Bielefeld},
country={Germany}}
\begin{abstract}
We will report on current progress in the understanding of the QCD phase diagram, including universal scaling in the chiral limit and the vicinity of the QCD critical point. In the latter case we will discuss the universal scaling of Lee-Yang zeros and their determination from multi-point Padé approximations to the baryon number density at imaginary chemical potentials. In particular, reported results include the critical phase transition temperature, the curvature of the critical and pseudo-critical transition temperature with respect to the chemical potential and the location of the QCD critical point. 
\end{abstract}

\end{frontmatter}

\section{Introduction}
Large efforts are made to improve and consolidate our knowledge on the QCD phase diagram. 
This includes large scale experimental heavy ion programs at the Relativistic Heavy Ion Collider (RHIC) in Brookhaven, NY and the Large Hadron Collider (LHC) in Genvea, Switzerland, as well as theoretical studies and numerical lattice QCD calculations (see, e.g. \cite{Bzdak:2019pkr, Pasztor:2024dpv}). 
A deep understanding of the QCD phase diagram, including in particular locations of phase transition points and strength and size of corresponding critical regions which are dominated by universal critical phenomena is mandatory to answer many pressing questions in cosmology, astrophysics and heavy ion phenomenology.  

The QCD phase diagram is largely determined by the chiral symmetry of the QCD Lagrangian. 
It is well known that the Lagrangian exhibits in the limit of $N_f$ massless quarks a symmetry under the independent rotations of left- and right handed Dirac spinors $U_L(N_f)\times U_R(N_f)$. 
This symmetry can also be expressed in terms of an unidirectional (vector) and an opposite (axial) rotations of the spinors  $U_V(N_f)\times U_A(N_f)$.
It is instructive to decompose the $U(N_f)$ groups into a $U(1)$ subgroup and the remaining coset $SU(N_f)/Z(N_f)$. 
While the $U_V(1)$ symmetry is responsible for the baryon number conservation, the $U_A(1)$ symmetry is broken in the vacuum by the axial anomaly and is strongly related to the topological properties of the theory \cite{Pisarski:1983ms, Pisarski:2024esv}. 
The remaining symmetry, i.e. $SU_V(N_f)\times SU_A(N_f)$ is broken spontaneously to $SU_V(N_f)$ and restored at high temperatures for $T\ge T_c$. 
In addition, the mass term in the Lagrangian also breaks the axial symmetry $SU_A(N_f)$ explicitly. 
For two light quark flavor $N_f=2$, we find that the direct product $[SU_V(2)/Z(2)]\times[SU_A(2)/Z(2)]$ is isomorphic to $O(4)$. 
The symmetry breaking pattern of the spontaneous symmetry breaking will thus be $O(4)\to O(3)$ and the relevant universality class of the transition is that of the 3d $O(4)$ symmetric spin model. 
In the chiral limit we thus expect a second order phase transition. 
However, here are two remarks in order: (i) it is still under debate weather or not the axial anomaly is effectively restored at $T=T_c$. If so, the symmetry breaking pattern will be $U_V(2)\times U_A(2)\to U_V(2)$, which might also alter the nature of the transition \cite{Pisarski:2024esv, Pelissetto:2013hqa}. (ii) Calculations we report on here are performed using the staggered discretization schema, which preserves only a sub-group of the chiral symmetry. Therefore we expect that the symmetry breaking pattern will be $O(2)\to Z(2)$ for calculations with finite lattice spacing $(a)$. In the limit $a\to 0$ we expect that the $O(4)$ symmetry is restored.  

\begin{figure*}
    \centering
    \includegraphics[width=0.99\textwidth]{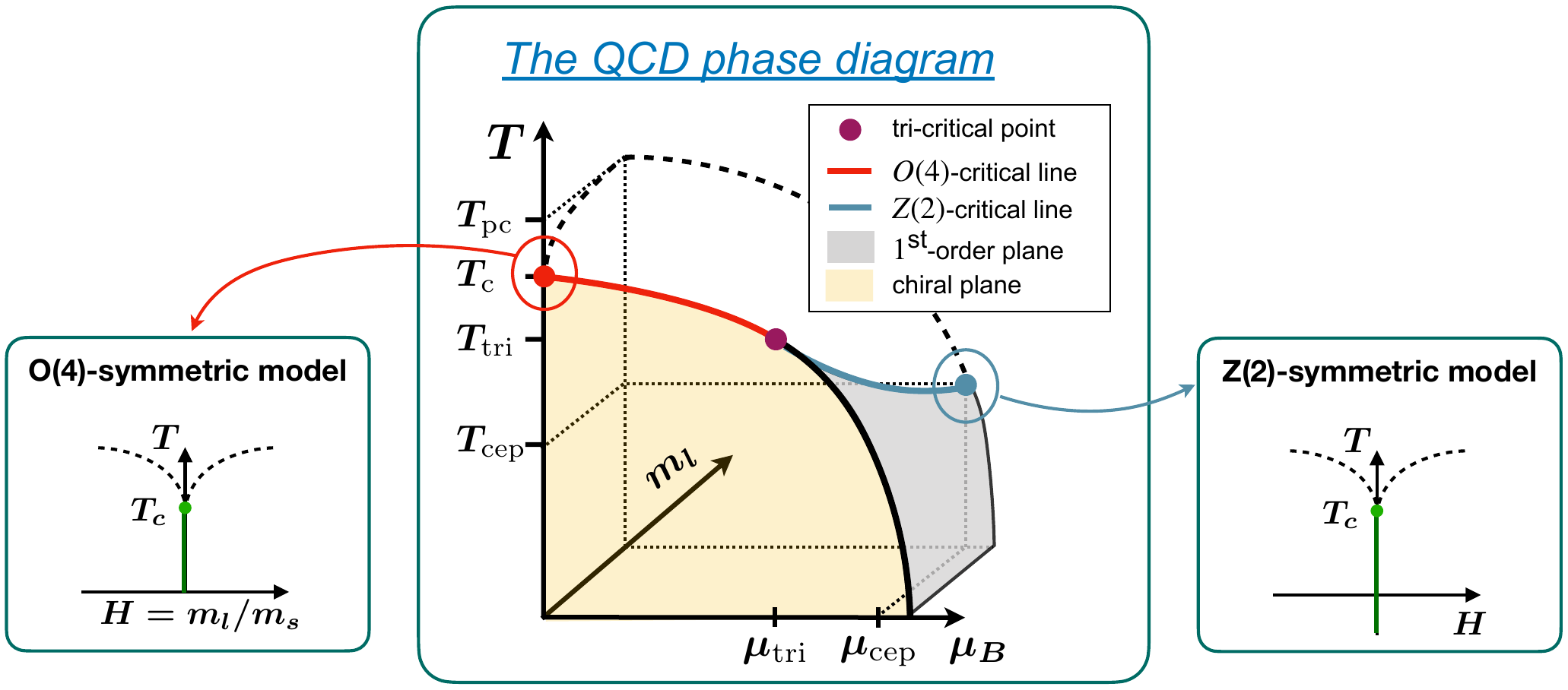}
    \caption{Schematic representation of the QCD phase diagram in the space of Temperature $T$, light quark mass $m_l$, and net baryon chemical potential $\mu_B$. 
    Shown are second order phase transition lines and first order planes. 
    Also indicated are the scaling regions around the chrial, and QCD critical points, which can be mapped to the universal O(4) and Z(2) symmetric spin models, respectively. }
    \label{fig:pdiag}
\end{figure*}

In the case of (2+1)-flavor, i.e. two degenerate light quarks and one heavier strange quark, the picture could be disturbed. 
The presence of the heavier quark might induce a first order transition at sufficiently small light quark masses \cite{Giacosa:2024orp}. However, recent lattice studies with (2+1) \cite{HotQCD:2019xnw, Ding:2024sux} and 3 \cite{Dini:2021hug} degenerate flavors of highly improved staggered quarks (HISQ) seem to exclude this scenario. 
Our current understanding of the (2+1)-flavor QCD phase diagram, spanned by temperature $(T)$, light quark mass $(m_l)$ and baryon chemical potential $(\mu_B)$ is summarized in Fig.~\ref{fig:pdiag}.  
Solid lines indicate phase transition lines, dashed lines indicate pseudo-critical lines, where the transition is a crossover. 
The second order $O(4)$ critical transition extends into the $\mu_B>0$ region until it turns first order at a tri-critical point. 
From here also a $Z(2)$ critical line branches off, which is connected with the QCD critical point at physical quark masses. 
In order to investigate the universal critical behaviour, it is our aim to map QCD parameters to the universal scaling fields in the chiral limit and at the QCD critical point.

\section{Universal scaling near the chiral critical point}
\begin{figure*}
    \centering
    \includegraphics[width=0.49\textwidth]{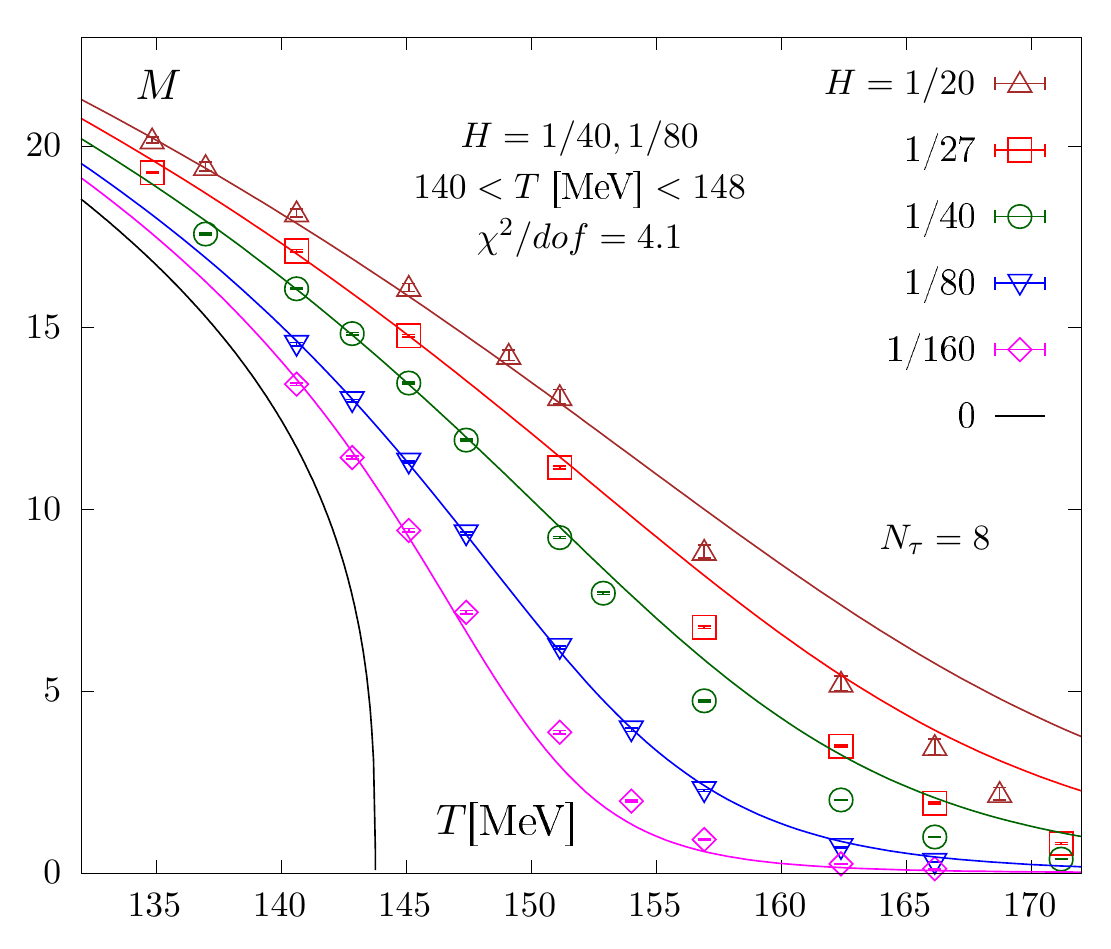}
    \includegraphics[width=0.49\textwidth]{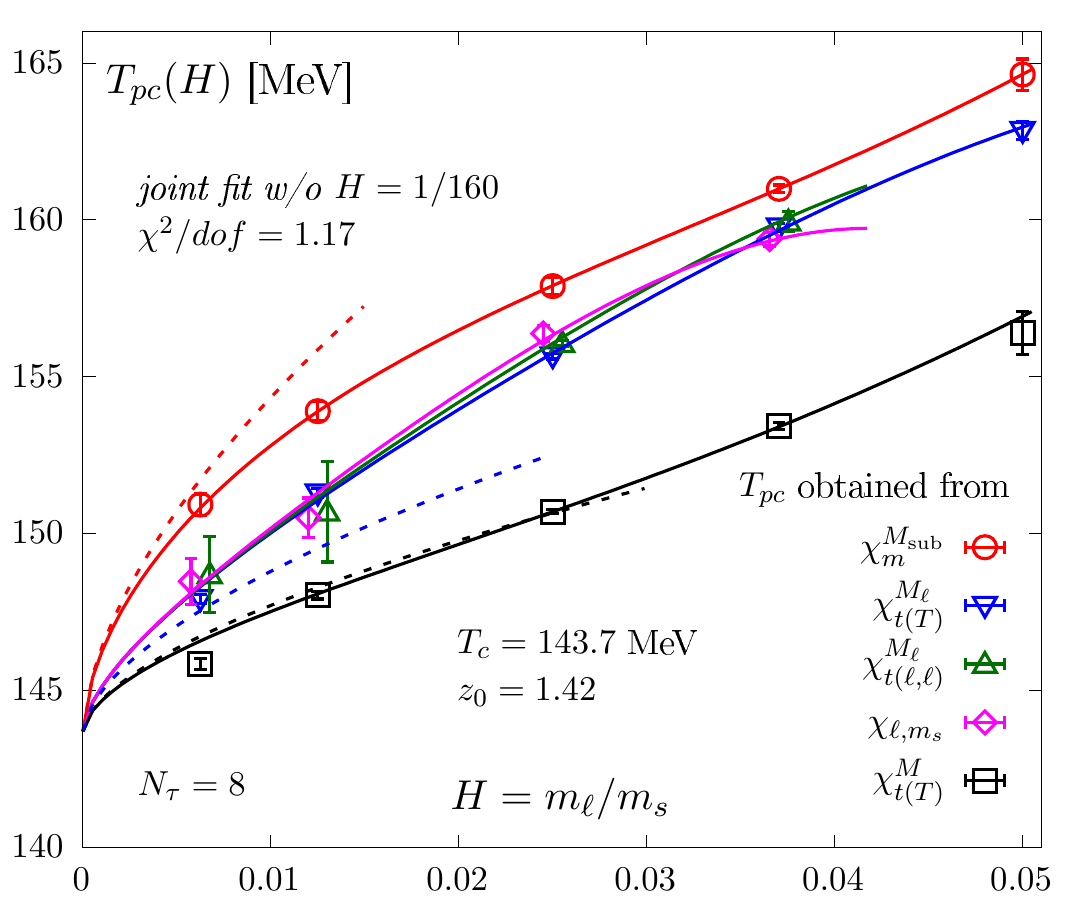}
    \caption{Left: Improved order parameter as function of the temperature for various quark mass ratios and fit to the magnetic equation of state. Right: pseudo-critical temperature obtained from different observables as function of the symmetry breaking field $H$ and fit to the expected universal critical behaviour including corrections to scaling. The dashed lines indicate the leading singular behaviour. The plots are taken from \cite{Ding:2024sux}.}
    \label{fig:chiral}
\end{figure*}

On the O(4) critical line, the symmetry breaking field $h$ is solely driven by the light quark mass. All other QCD parameters do not break chiral symmetry and will thus contribute to the temperature like scaling field $t$. We assume the following ansatz for the mapping to the scaling fields \cite{Ding:2024sux}, 
\begin{align}
     h&=\frac{1}{h_0}H=\frac{1}{h_0}\frac{m_l}{m_s} \\
    t&=\frac{1}{t_0}\left(\Delta T + \kappa_2^l\hat\mu_l^2+ \kappa_2^s\hat\mu_s^2 +2\kappa_{11}^{ls}\hat\mu_l\hat\mu_s\right)\,.
\end{align}
Here we define $\Delta T=T/T_c-1$, and $m_l,m_s$ are the light and strange quark masses, the latter is being kept fixed to its physical value. The light and strange chemical potentials are determined as $\mu_l$ and $\mu_s$, we further define $\hat\mu_x=\mu_x/T$ for $x\in\{l,s,B\}$.
The critical temperature $T_c$, the normalization constants $h_0$, $t_0$ and the couplings $\kappa_2^{l}$, $\kappa_2^{s}$, $\kappa_{11}^{ls}$, are none-universal constants that need to be determined from fits to the lattice data. 
In particular we calculate the order parameter $(M_l)$, the magnetic susceptibility $(\chi_l)$, defined as 
\begin{align}
    M_l&=\frac{m_s}{f_K^4}\frac{T}{V}\frac{\partial \ln Z}{\partial m_l}\\
    \chi_{l}&=m_s\frac{\partial }{\partial m_l}M_l\,.
\end{align}
Note that each derivative $\partial/\partial m_l$ is accompanied by a factor $m_s$, which is a consequence by the definition of the symmetry breaking field $H=m_l/m_s$, which makes $H$ dimensionless but also removes multiplicative UV divergences. Although these two observables are sufficient to investigate universal critical behaviour, we can construct an improved order parameter $M$ that is free of additive UV divergences and the leading regular contribution proportional to $H$, by defining
\begin{equation}
    M=M_l-H\chi_l\,.
\end{equation}
As $M$ is constructed from $H$ derivatives of $\log Z$, we can also construct an equation of state for this order parameter \cite{Kotov:2021rah}, given as 
\begin{equation}
    M=h^{1/\delta}(f_G(z)-f_\chi(z))\,
\end{equation}
where we define $z=t/h^{1/\beta\delta}$ as scaling variable. 
In Fig.~\ref{fig:chiral} (left) we show a fit of the lattice data for $M$ to the equation of state, where we use the $O(2)$ scaling functions determined in Ref.~\cite{Karsch:2023pga}.
The lattice data stems from calculations on $N_\sigma^3\times N_\tau$ lattices, with (2+1)-flavor of highly improved staggered quarks (HISQ). The spacial extent $N_\sigma$ was adjusted to the quark mass ratio, in order to keep finite size effects small. 
It varies between $N_\sigma=32$ for $H=1/27$ and $N_\sigma=56$ for $H=1/160$. 
In fact, we find that at $H=1/160$, despite the larger spacial extent,  the finite size effects are more pronounced than for the larger masses. For this reason we reject this quark mass ratio from the fit. 
The fit yields a critical temperature of $T_c=143.9(5)$ MeV for $N_\tau=8$, which is in agreement with our previous determination of $T_c$ from $\chi_l$, which leads to a continuum extrapolated result of $T_c=132^{+3}_{-6}$ MeV \cite{HotQCD:2019xnw}.

A determination of the peak positions of various mixed susceptibilities, including $\partial M_l/\partial T$, $\partial M_l/\partial \mu_l^2$, $\partial M/\partial T$, are shown in Fig.~\ref{fig:chiral} (right) \cite{Ding:2024sux}. 
A joint fit of the data to the universal scaling is indicated by the solid lines. 
The fit includes corrections to scaling terms. 
The asymptotic behavior for small $H$ is shown by the dashed lines, which corresponts to lines of constant $z$-values. 
The fit yields yet another independent determination of $T_c$ for $N_\tau=8$, which is $T_c=143.9(5)$ MeV. 
Again, this is fully consistent with our previous continuum estimate \cite{HotQCD:2019xnw}. Interestingly, we can determine the curvature coefficients $\kappa_2^l$, $\kappa_2^s$ and $\kappa_{11}^{ls}$ by taking appropriate ratios of the mixed susceptibilities. 
This is possible since the universal scaling of, e.g. $\partial M_l/\partial \mu_l^2$ and $\partial M_l/\partial T$ differ only by a factor of of $\kappa_2^l$. 
We can convert the curvature coefficients into the phenomenologically more relevant hadronic basis and obtain $\kappa_2^{B,\mu_S=0}=0.015(1)>\kappa_2^{Bn_S=0}=0.893(35) \kappa_2^{B,\mu_S=0}$ \cite{Ding:2024sux}, where coefficient are calculated at constant $\mu_S=0$ and $n_S=0$, respectively.  

\section{The QCD critical end-point}
A study of the universal scaling near the QCD critical end-point (CEP) at physical quark masses and $\mu_B>0$ is much more difficult. This is not only due to the infamous sign problem in QCD that prevents direct lattice QCD calculations at $\mu_B>0$, but also because the scaling directions are not known a priori. Here we assume a frequently used linear mixing ansatz for the scaling fields \cite{Rehr:1973zz, Nonaka:2004pg}
\begin{eqnarray}
t&=&\alpha_t(T-T_{cep})+\beta_t(\mu_B-\mu_{cep})\,, \nonumber \\
h&=&\alpha_h(T-T_{cep})+\beta_h(\mu_B-\mu_{cep})\,,
\label{eq:scaling_fields}
\end{eqnarray}
where $\alpha_t,\alpha_h, \beta_t$ and $\beta_h$ are the mixing parameters. 
One can easely verify that the ratio $-\beta_h/\alpha_h$ defines the slope of the first order line at the QCD critical point. 
The data we consider here are derivatives of $\ln Z$ with respect to $\hat\mu_B$ at imaginary chemical potential $\mu_B=i\theta_B$ and physical quark masses. 
We define a set of observables as 
\begin{equation}
\chi_n^B(T,V,\mu_B)=\left(\frac{\partial}{\partial\hat\mu_B}\right)^n\frac{\ln Z(T,V,\mu_B)}{VT^3}\,.
\end{equation}
We use diagonal rational approximations to the data $\chi_1^B$ of the form 
\begin{equation}
    R^n_n(\hat{\mu}_B)=\frac{\sum_{j=0}^n a_j\hat{\mu}_B^j}{1+\sum_{j=1}^n b_j\hat{\mu_B}^j}\,.
    \label{eq:ratapprox}
\end{equation}
For the analysis we use $n=3,4,5$. 
We determine the free parameter $a_j,b_j$ by solving a system of linear equations involving also the data for $\chi_2^B$ as well as the derivative $\text{d}R_n^n/\text{d}\hat\mu_B$ at a number of imaginary simulation points in the range $\hat\mu_B\in[0,i\pi]$. We refer to this approximation as multi-point Padé. We further identify the closest pole of the approximation in the complex chemical potential plane with the Lee-Yang edge. This strategy was successfully tested in the case of the Roberge-Weiss transition in QCD \cite{Dimopoulos:2021vrk} and the 2d-Ising Model \cite{Singh:2023bog}, where in both cases known results could be recovered. We note that we take into account cancellation of zeros in the numerator and zeros in the denominator of Eq.~(\ref{eq:ratapprox}), which might be inexact due to numerical errors. 

\begin{figure}
    \centering
    \includegraphics[width=0.48\textwidth]{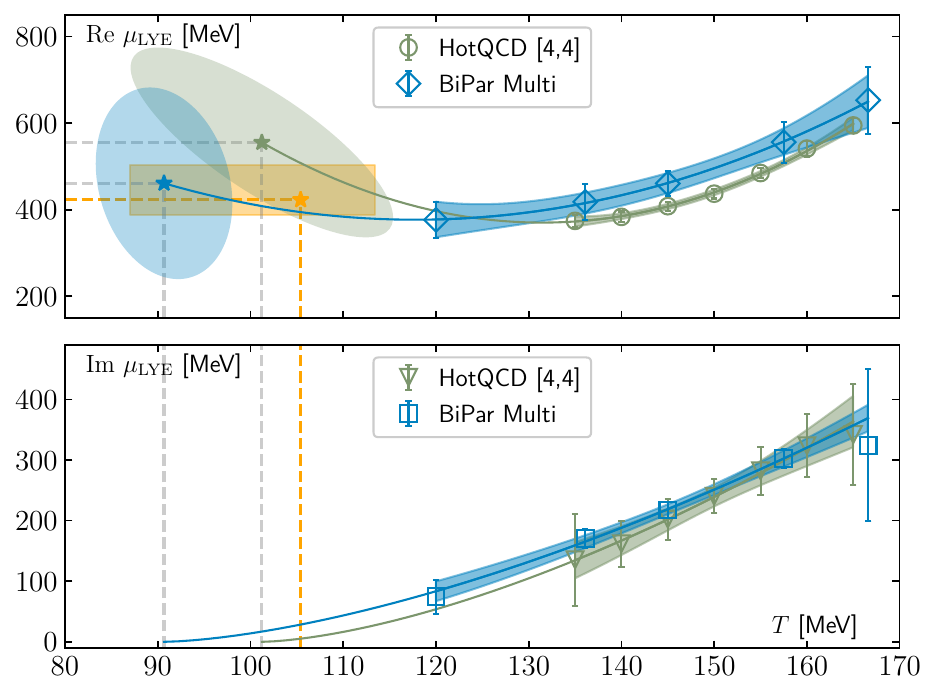}
    \caption{Scaling fits for the Lee-Yang singularities related to the QCD critical point. Green data come from a [4,4] Pad\'e from Ref.~\cite{Bollweg:2022rps}.
    Blue data come from the multi-point Padé \cite{Clarke:2024ugt}. 
    {\it Top}: Scaling of the real part. {\it Bottom}: Scaling of the
    imaginary part. The ellipses shown in the top panel represent the 68\% confidence region deduced from the covariance matrix of the fit. The orange box indicates the AIC weighted estimate \cite{Clarke:2024ugt}}
    \label{fig:CEP}
\end{figure}

The Lee-Yang edge exhibits a well defined universal scaling behaviour, which is expressed by its universal constant location in the complex plane of the scaling variable $z$, i.e. $t/h^{1/\beta\delta}=z_{LY}$ \cite{Connelly:2020gwa}. 
Moreover, the Lee-Yang theorem \cite{Yang:1952be} states that for $t>0$ we expect the Lee-Yang edge at imaginary $h_{LY}$. 
Together with the ansatz for the scaling fields Eq.~(\ref{eq:scaling_fields}), we can now derive the temperature behaviour of the Lee-Yang edge in the complex chemical potential plane. 
We find that $\Im[\mu_{LY}]
\sim(T-T_{cep})^{\beta\delta}$, for $T\searrow T_{cep}$, while $\Re[\mu_{LY}]=\mu_{cep}-\beta_h/\alpha_h(T-T_{cep})+\mathcal{O}(T^2)$ \cite{Stephanov:2006dn}. 
A fit to the Lee-Yang scaling is shown in Fig.~\ref{fig:CEP} for $N_\tau=6$ lattices. 
The results are compared with the [4,4]-Pade results constructed out of the eights order Taylor expansion about $\mu_B=0$, for $N_\tau=8$ \cite{Bollweg:2022rps}. Both calculations give similar results. 
The location of the critical point is determined to be $(T_{cep},\mu_{cep})=(105^{+8}_{-18}, 422^{+80}_{-35})$ for $N_\tau=6$. For more details on the analysis see \cite{Clarke:2024ugt}. We note that this result is not yet continuum extrapolated. Systematic errors that are included here through a variation of the order of the multi-point Padé will be further investigated in future studies.  

\section*{Acknowledgments}
CS acknowledges V. Skokov and all members of the HotQCD and Bielefeld-Parma Collaboration for valuable discussions. 
This work was supported (i) by the European Union’s Horizon 2020 research and innovation program under the Marie Sklodowska-Curie Grant Agreement No. H2020-MSCAITN-2018-813942 (EuroPLEx), (ii) by The Deutsche Forschungsgemeinschaft (DFG, German Research Foundation) - Project No. 315477589-TRR 211 and the PUNCH4NFDI consortium supported by the Deutsche Forschungsgemeinschaft (DFG, German Research Foundation)
with grand 460248186 (PUNCH4NFDI).

\bibliography{mybib}{}
\bibliographystyle{elsarticle-num}
\end{document}